# Quantum Entanglement in Nitrosyl Iron Complexes

S. M. Aldoshin, E. B. Feldman, and M. A. Yurishchev

*Institute for Problems of Chemical Physics, Russian Academy of Sciences, Chernogolovka, Moscow oblast, 142432 Russia*
*e-mail: efeldman@icp.ac.ru*

**Abstract**—Recent magnetic susceptibility measurements for polycrystalline samples of binuclear nitrosyl iron complexes, $[Fe_2(C_3H_3N_2S)_2(NO)_4]$ (I) and $[Fe_2(SC_3H_5N_2)_2(NO)_4]$ (II), suggest that quantum-mechanical entanglement of the spin degrees of freedom exists in these compounds. Entanglement $E$ exists below the temperature $T_E$ that we have estimated for complexes I and II to be 80–90 and 110–120 K, respectively. Using an expression of entanglement in terms of magnetic susceptibility for a Heisenberg dimer, we find the temperature dependence of the entanglement for complex II. Having arisen at the temperature $T_E$, the entanglement increases monotonically with decreasing temperature and reaches 90–95% in this complex at $T = 25$ K, when the side effects are still small.

## 1. INTRODUCTION

Entanglement is one of the most intriguing quantum-mechanical phenomena. A system of two spins in a state with the wave function

$$|\psi\rangle = (|\uparrow\downarrow\rangle + |\downarrow\uparrow\rangle)/\sqrt{2}$$

can serve as an example where this phenomenon manifests itself. This function, which describes a coherent superposition of qubits, cannot be represented as a product of the wave functions of the system's individual constituents (the state is not separable). On the other hand, because of this property, the property of entanglement, measuring the state of one particle allows the state of the second particle to be instantly reduced no matter how far or close it is from the first one.

At present, entanglement and related possibilities of quantum calculations, cryptography, teleportation, etc. are investigated not only theoretically but also experimentally. Moreover, there are real prerequisites for using these unique possibilities of quantum mechanics in practice already now.

The literature on this subject matter is very extensive and diverse; to be specific, we will point out, for example, the reviews and books [1–5] as well as the sites www.qubit.org and xxx.arxiv.ru.

Important relationships that allow predictions about the existence of entanglement in systems to be made using such experimentally measurable characteristics as the correlation functions, internal energy, and magnetic susceptibility have been established in recent years [6–9] (see also the review [10] and the dissertation [11]).

These theoretical results have opened a possibility of determining the temperature $T_E$ at which entanglement arises in actual materials. Brukner et al. [12] were the first to determine this temperature. They showed that $T_E \approx 5$ K in paramagnetic $Cu(NO_3)_2 \cdot 2.5H_2O$ and $Cu(NO_3)_2 \cdot 2.5D_2O$ crystals. In another quite recent paper, Souza et al. [13] presented evidence suggesting that quantum entanglement arises in $Na_2Cu_5Si_4O_{14}$ crystals containing chains of pentanuclear copper spin clusters at temperatures below $T_E \approx 200$–240 K.

In this paper, we consider the recently synthesized nitrosyl iron complexes (NICs) I [14] and II [15].

Their physical properties have been studied by various methods, including X-ray analysis, Mössbauer and infrared spectroscopy. In addition (this is particularly important for the subsequent discussion), the temperature dependences of the magnetic susceptibilities were obtained for these NICs using SQUID magnetometers [14, 15].

In our paper, for the above materials, we not only determine the temperature at which intramolecular entanglement arises but also derive the temperature dependence of the entanglement for NIC II using the Wootters formula (see below).

In addition to this introduction, our paper contains the following. In Section 2, we describe the structure and physical parameters of NICs. In Section 3, we construct a model and derive formulas for calculating the quantum entanglement from the magnetic susceptibility. Our results are discussed and interpreted in Section 4. In Section 5, we briefly summarize our results and give conclusions.

## 2. THE DIMER MAGNETIC STRUCTURE OF BINUCLEAR NICs

Nitrosyl complexes are the carriers of nitrogen monoxide (NO), which acts as a signal molecule in a number of metabolic and physiological processes that take place in biological systems and organisms, including humans. NICs were discovered in living tissues in the 1960s [16]. They were called "2.03 complexes" by the characteristic EPR signal with $g = 2.03$.

In living tissues, NICs exist in mononuclear and binuclear forms, respectively, with one and two iron ions in the molecule. Natural binuclear NICs are unstable and their study by physical methods is very limited. Methods for the synthesis of several stable artificial NICs have been developed in the 2000s.

They allow one to indirectly determine (model) the characteristics of natural NICs and open a possibility for creating preparations with the needed medico-biological and pharmacological properties [17].

The structure of the binuclear NICs of interest to us is

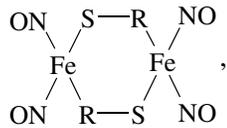

where R is imidazole-2-thiolate (for complex I) or imidazolidine-2-thiolate (for complex II). The ligands R are five-membered heterocycles that consist of three carbon atoms and two nitrogen atoms separated by carbon.

X-ray analysis, Mössbauer spectroscopy, and magnetic susceptibility measurements yield the following results [14, 15, 18].

The magnetoactive centers are formed by iron ions each of which is linked with two nitrosyl groups.

The spin of an individual center is $S = 1/2$. (The formation mechanism of a magnetoactive center from one Fe and two NO groups was considered for a mononuclear NIC as an example in [19].)

The molecules have a centrosymmetric structure in which the two iron atoms are connected by two S–C–N bridges, where C–N is a fragment of the R heterocycle. As a result, a paramagnetic dimer is formed. The interaction in it obeys the Heisenberg law and is antiferromagnetic in nature. The Fe–Fe separation in complexes I and II are 0.4102 and 0.4030 nm, respectively. (A smaller separation increases the interaction force in the magnetic dimer of complex II.)

The crystals of the NICs under discussion have a layered structure. Intermolecular contacts of the sulfur atom and the H–N fragment in the ring of the ligand R exist inside the layer between the molecules in one direction. The presence of such contacts results in the existence of interdimer interactions, but they are weak due to the large extent of these superexchange couplings. The molecules from different layers are oriented to one another by the NO groups. The layers are coupled only by electrostatic forces, which also contributes to an isolation of the magnetic dimers from each other.

As a result, our studies lead us to the following conclusions. First, the magnetic structure of the binuclear NICs I and II comprises isolated dimers. This distinguishes the materials under discussion among other materials that exhibit quasi-dimer magnetic properties (see the reviews [20, 21] and references therein). Second, magnetic susceptibility measurements suggest that complexes I and II are antiferromagnetic and the interactions in the dimers are Heisenberg ones.

## 3. THE MODEL AND CALCULATION OF ENTANGLEMENT

The Hamiltonian of a Heisenberg spin magnetic dimer is

$$\mathcal{H} = -\frac{1}{2}J\sigma_1\sigma_2. \qquad (1)$$

Here, $J$ is the exchange coupling constant, $\sigma_1 = \sigma \otimes e$, and $\sigma_2 = e \otimes \sigma$, where $e$ is a $2 \times 2$ unit matrix and $\sigma = (\sigma_x, \sigma_y, \sigma_z)$ is the vector of Pauli matrices

$$\sigma_x = \begin{pmatrix} 0 & 1 \\ 1 & 0 \end{pmatrix}, \quad \sigma_y = \begin{pmatrix} 0 & -i \\ i & 0 \end{pmatrix},$$
$$\sigma_z = \begin{pmatrix} 1 & 0 \\ 0 & -1 \end{pmatrix}. \qquad (2)$$

The magnetic moment components for the dimer are

$$M_\nu = \frac{1}{2}g_\nu\mu_B(\sigma_1^\nu + \sigma_2^\nu), \quad \nu = x, y, z. \qquad (3)$$

Here, $g_\nu$ are the components of the $g$ factor and $\mu_B$ is the Bohr magneton.

The magnetic susceptibility of a mole of dimers (1)–(3) satisfies the Bleaney–Bowers equation [22, 23]

$$\chi(T) = \frac{2N_A g^2 \mu_B^2}{k_B T(3 + \exp(-2J/k_B T))}, \qquad (4)$$

where $N_A$ is the Avogadro number, $k_B$ is the Boltzmann constant, and $T$ is the temperature; $g$ is the corresponding component of the $g$ factor if the measurements are made on a single crystal or

$$g^2 = (g_x^2 + g_y^2 + g_z^2)/3 \qquad (5)$$

if the measurements are made on a polycrystalline sample. We will need the Bleaney–Bowers equation to establish the relationship between entanglement and magnetic susceptibility.

The entanglement in the formula

$$E = -\frac{1+\sqrt{1-C^2}}{2}\log_2\left(\frac{1+\sqrt{1-C^2}}{2}\right) \\ -\frac{1-\sqrt{1-C^2}}{2}\log_2\left(\frac{1-\sqrt{1-C^2}}{2}\right) \quad (6)$$

is expressed in terms of the so-called concurrence $C$ [24–26]. The density matrix must be known to calculate it.

The density matrix of a system in thermal equilibrium has the Gibbs form

$$\rho = \frac{1}{Z}\exp\left(-\frac{\mathcal{H}}{k_B T}\right), \quad (7)$$

where $Z$ is the partition function

$$Z = \text{Tr}\exp\left(-\frac{\mathcal{H}}{k_B T}\right). \quad (8)$$

It is easy to verify that

$$\sigma_1\sigma_2 = \begin{pmatrix} 1 & 0 & 0 & 0 \\ 0 & -1 & 2 & 0 \\ 0 & 2 & -1 & 0 \\ 0 & 0 & 0 & 1 \end{pmatrix}. \quad (9)$$

As a result, the density matrix of a Heisenberg dimer (1) is

$$\rho(T) = \frac{1}{Z}\begin{pmatrix} e^K & & & \\ & e^{-K}\cosh 2K & e^{-K}\sinh 2K & \\ & e^{-K}\sinh 2K & e^{-K}\cosh 2K & \\ & & & e^K \end{pmatrix}, \quad (10)$$

where

$$Z = 3e^K + e^{-3K} \quad (11)$$

and $K = J/2k_B T$.

In the case of a density matrix with a block-diagonal structure of the form

$$\rho = \begin{pmatrix} u & & & \\ & x_1 & w & \\ & w^* & x_2 & \\ & & & v \end{pmatrix}, \quad (12)$$

the following simple formula [27] (see also [4, pp. 49, 55]) is used to calculate the concurrence:

$$C = 2\max\{|w| - \sqrt{uv}, 0\}. \quad (13)$$

Our density matrix (10) has form (12).

Equation (13) is a special case of the well-known Wootters formula [25, 26] (see also [4, p. 48]) that allows the pair concurrence between particles with spins $S = 1/2$ in a system with a density matrix ρ of arbitrary structure to be calculated.

Using Eqs. (10)–(13), we can easily find that the concurrence is identically equal to zero in dimer (1) with a ferromagnetic coupling, when $J \geq 0$. In view of (6), entanglement is also absent at all temperatures: $E \equiv 0 \; \forall T$.

In contrast, according to Eqs. (10)–(13), the concurrence for an antiferromagnetic coupling in a Heisenberg dimer ($J < 0$) is [28, 29] (see also [4, p. 50])

$$C(T) = \frac{1 - 3\exp(-2|J|/k_B T)}{1 + 3\exp(-2|J|/k_B T)}, \quad T < T_E, \\ C(T) = 0, \quad T \geq T_E, \quad (14)$$

where

$$T_E = \frac{2}{\ln 3}|J|/k_B. \quad (15)$$

Note that the Wootters formula from which Eq. (14) was derived was proved rigorously directly from the definition of entanglement via the von Neumann information entropy [26]. Therein lies its advantage over other, formal measures of entanglement (see [10] and references therein). Note also that the existence of entanglement in the system may lead to a breakdown of the Bell inequalities [29].

Taking into account the Bleaney–Bowers equation (4), we obtain the following expression for the concurrence of an antiferromagnetic Heisenberg dimer (1) at temperatures $T < T_E$ from (14):

$$C(T) = 1 - \frac{3}{2}\frac{\chi(T)}{\chi^{\text{Curie}}(T)}, \quad (16)$$

where

$$\chi^{\text{Curie}}(T) = \frac{N_A g^2 \mu_B^2}{2k_B T} \quad (17)$$

is the Curie law for two spins with $S = 1/2$ (the Bleaney–Bowers equation (4) at high temperatures, when it may be assumed that $J = 0$). These relations, along with Eq. (6), allow the quantum entanglement to be determined from the experimentally measurable magnetic susceptibility of a system of Heisenberg dimers.

## 4. RESULTS AND DISCUSSION

The magnetic susceptibility of an antiferromagnetic dimer (4) as a function of the temperature has a maximum with coordinates

$$\frac{k_B T_{\max}^\chi}{|J|} = \frac{2}{1 + W(3/e)} = 1.2472..., \quad (18)$$

$$\frac{|J|\chi_{\max}}{N_A g^2 \mu_B^2} = \frac{1}{3} W(3/e) = 0.2011\ldots . \quad (19)$$

Here, $W(x)$ is the Lambert function [30] defined by the relation $We^W = x$. This function under the name Lambert$W(x)$ was included in the Maple package.

We find from (15) and (18) that

$$\frac{T_E}{T_{\max}^\chi} = \frac{1 + W(3/e)}{\ln 3} = 1.4596\ldots . \quad (20)$$

Thus, quantum entanglement of the spin degrees of freedom arises at a temperature that is almost a factor of 1.5 higher than the temperature of the magnetic susceptibility maximum. This is favorable for the determination of $T_E$ and $E(T)$ from experimental data.

In accordance with (16), entanglement in dimer (1) exists when its susceptibility is

$$\chi(T) < \frac{2}{3}\chi^{\text{Curie}}(T). \quad (21)$$

This inequality is in complete agreement with a more general inseparability criterion—the condition for the emergence of entanglement in a system of $N$ particles with spins $S$ [9, 11]:

$$\chi_p(T) < \frac{Ng^2\mu_B^2 S}{3k_B T}, \quad (22)$$

where $\chi_p$ is the susceptibility averaged over the spatial directions (the susceptibility of a polycrystalline material).

We will represent Eq. (22) as

$$\chi_p(T) < \frac{1}{1+S}\chi^{\text{Curie}}(T), \quad (23)$$

where

$$\chi^{\text{Curie}}(T) = \frac{nN_A g^2 \mu_B^2 S(S+1)}{3k_B T} \quad (24)$$

is the Curie law for a mole of $n$-nuclear clusters consisting of spins $S$. (In a dimer, $n = 2$.)

The right-hand sides of inequalities (21) and (23) are deformed Curie laws. The entanglement in the system can be determined precisely due to these deformations (because of the additional renormalization coefficients).

Note that criterion (23), just as (21), allows only the temperature $T_e$ but not the entanglement itself to be found. However, for dimer compounds, and therein lies their advantage, we can both determine $T_E$ and reproduce the temperature dependence of the entanglement of various materials with the help of Eqs. (6), (16), and (17) using experimental data on the susceptibility.

Let us turn to the experimental data. Figure 1 shows the behavior of the initial magnetic susceptibility for complex I [14]. The actual material with NIC I contains a small amount of impurity (~2.3% [14]) whose contribution obeys the Curie–Weiss law. This contribution is small when $T \longrightarrow 0$ (in Fig. 1, it manifests itself as a rise in susceptibility near zero temperature). However, the contribution from the impurity is relatively small at $T > T_{\max}^\chi$ and we will neglect it when estimating the temperature $T_E$ for NIC I.

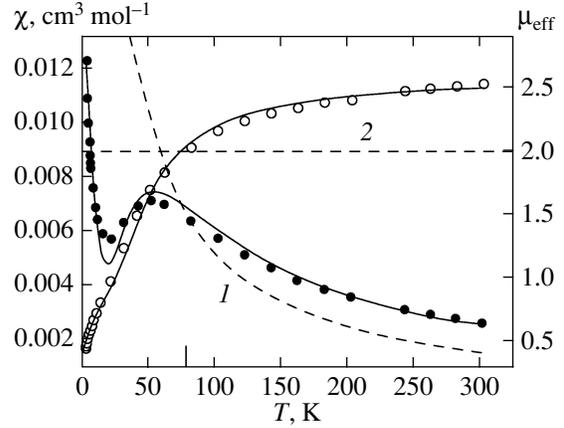

**Fig. 1.** Temperature dependences of the magnetic susceptibility (●) and effective magnetic moment (○) for NIC I. We took this figure from [14] and supplemented it, as explained in the text, by the dashed lines *1* and *2*.

The dashed line *1* in Fig. 1 indicates a deformed Curie hyperbola (the right-hand side of inequality (21)), given that $g = 2$ in NIC [14]. We see from the figure that this hyperbola crosses the experimental data points (filled circles) at $T_E \approx 80$ K (along the horizontal axis, this temperature is marked by the longer bar). On the other hand, as was pointed out in [18], the magnetic susceptibility of NIC I under discussion passes through a maximum at $T_{\max}^\chi = 63$ K. Consequently, in accordance with Eq. (20), the entanglement temperature should be $T_E \approx 90$ K. Taking into account both estimates, we conclude that entanglement arises in complex I at temperatures $T < 80–90$ K.

In magnetochemistry, it is common practice to represent the same experimental data, along with the curves $\chi(T)$ (and often even instead of them), as the temperature dependences of an effective magnetic moment

$$\mu_{\text{eff}}(T) = \frac{1}{\mu_B}\left(\frac{3k_B}{N_A}T\chi\right)^{1/2}. \quad (25)$$

Using (24), criterion (23) can then be written in the form

$$\mu_{\text{eff}}(T) < g\sqrt{nS}. \quad (26)$$

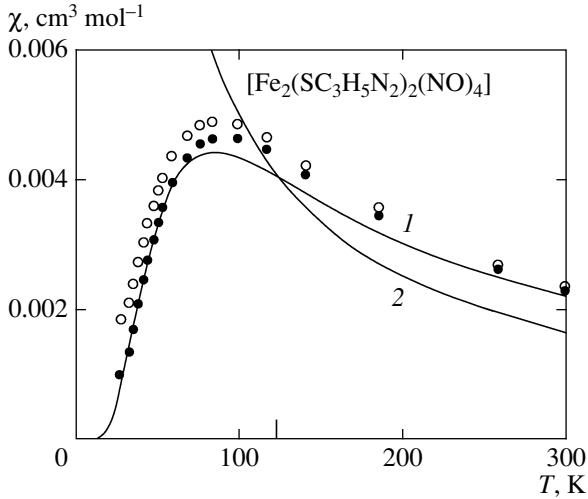

**Fig. 2.** Behavior of the magnetic susceptibility for NIC II with (open circles) and without (filled circles) impurities. Curve *1* represents the theoretical Bleaney–Bowers dependence with $J/k_B = -68$ K and $g = 2$; curve *2* represents the dependences $(2/3)\chi^{Curie}(T)$ with $g = 2$.

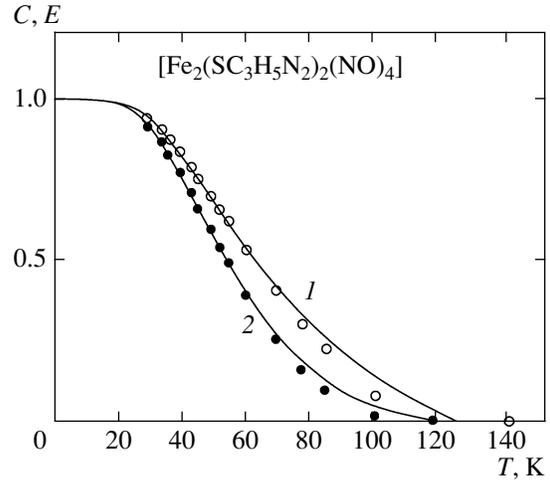

**Fig. 3.** Temperature dependences of the concurrence (open circles) and entanglement (filled circles) for NIC II. The solid curves *1* and *2* represent the theoretical dependences for $C$ and $E$, respectively.

It becomes very easy to use the criterion for the presence or absence of entanglement: it is only necessary to draw a horizontal straight line at the height $g\sqrt{nS}$ and check whether or not it intersects the curve $\mu_{eff}(T)$ and, if it does, determine the point of intersection.

In the combined Fig. 1, the open circles represent the values of $\mu_{eff}(T)$ for complex I. We drew the horizontal dashed straight line *2* in this figure at the level of $g\sqrt{nS} = 2$ (since $g = 2$, $n = 2$, and $S = 1/2$). We see that the abscissa of the point of intersection agrees with the temperature $T_E$ found above.

Let us now discuss the experimental data for NIC II published in [15]. In general, the behavior of the magnetic susceptibility here is similar to that for NIC I (see Fig. 2). The samples were purer; the amount of impurity did not exceed 1.7%. For a more careful analysis, we subtracted the contribution of the impurity and obtained $\chi(T)$ for NIC II proper (the filled circles in Fig. 2). In the same Fig. 2, curve *1* indicates the Bleaney–Bowers dependence (4) with the parameters found in [15]: $J/k_B = -68$ K and $g = 2$. As we see from Fig. 2, the deformed Curie curve crosses the filled circles at $T_E \approx 110$ K and the theoretical fit to the experimental data at $T_E \approx 120$ K. According to [18], the magnetic susceptibility of complex II with impurities under consideration passes through a maximum at a temperature of 83 K. Using Eq. (20), we find that $T_E \approx 121$ K. Thus, the temperature at which entanglement arises in NIC II is $T_E \approx 110$–120 K.

Finally, Fig. 3 presents the temperature dependences for the concurrence (open circles) and entanglement (filled circles) recalculated using Eqs. (6), (16), and (17) from experimental data on the susceptibility of complex II. The solid curves indicate the theoretical dependences $C(T)$ and $E(T)$ derived using the Bleaney–Bowers equation (4) with the above parameters $J/k_B$ and $g$ for this complex. We see from the figure that the degree of entanglement in the complex in 90–95% at $T = 25$ K.

The temperature behavior of the entanglement can be physically interpreted as follows. The Heisenberg dimer (1) has two energy levels, $3J/2$ and $-J/2$. They are separated by the energy gap $\Delta = 2|J|$. The first level is a singlet with the wave function

$$|\psi_0\rangle = \frac{1}{\sqrt{2}}(|\uparrow\downarrow\rangle - |\downarrow\uparrow\rangle) = \frac{1}{\sqrt{2}}\begin{pmatrix} 0 \\ 1 \\ -1 \\ 0 \end{pmatrix}. \quad (27)$$

The second level is threefold degenerate with the wave functions

$$|\psi_1\rangle = |\uparrow\uparrow\rangle = \begin{pmatrix} 1 \\ 0 \\ 0 \\ 0 \end{pmatrix}, \quad (28)$$

$$|\psi_2\rangle = |\downarrow\downarrow\rangle = \begin{pmatrix} 0 \\ 0 \\ 0 \\ 1 \end{pmatrix}, \quad (29)$$

$$|\psi_3\rangle = \frac{1}{\sqrt{2}}(|\uparrow\downarrow\rangle + |\downarrow\uparrow\rangle) = \frac{1}{\sqrt{2}}\begin{pmatrix} 0 \\ 1 \\ 1 \\ 0 \end{pmatrix}. \quad (30)$$

(The unit vectors are ordered as $|\uparrow\uparrow\rangle$, $|\uparrow\downarrow\rangle$, $|\downarrow\uparrow\rangle$, and $|\downarrow\downarrow\rangle$.)

If $J < 0$, then the singlet turns out to be the lower level. Passing to the limit $K \longrightarrow -\infty$ in Eqs. (10) and (11), we find that the density matrix at $T = 0$ is

$$\rho(0) = \frac{1}{2}\begin{pmatrix} 0 & & & \\ & 1 & -1 & \\ & -1 & 1 & \\ & & & 0 \end{pmatrix} = |\psi_0\rangle\langle\psi_0|. \quad (31)$$

Since this state is pure and maximally entangled, $E = 1$ in an antiferromagnetic Heisenberg dimer at absolute zero.

On the other hand, in the limit of an infinitely high temperature, when the spins behave as independent particles, the density matrix (10) transforms to

$$\rho(\infty) = \frac{1}{4}\begin{pmatrix} 1 & & & \\ & 1 & & \\ & & 1 & \\ & & & 1 \end{pmatrix} = \frac{1}{4}\sum_{i=0}^{3}|\psi_i\rangle\langle\psi_i|. \quad (32)$$

The particle entanglement in this maximally mixed, but, obviously, factorizable state is zero.

At the temperature $T_E$, the density matrix of an antiferromagnetic dimer is

$$\rho(T_E) = \frac{1}{6}\begin{pmatrix} 1 & & & \\ & 2 & -1 & \\ & -1 & 2 & \\ & & & 1 \end{pmatrix} \quad (33)$$

$$= \frac{1}{2}|\psi_0\rangle\langle\psi_0| + \frac{1}{6}\sum_{i=1}^{3}|\psi_i\rangle\langle\psi_i|.$$

In Appendix, we show through direct calculations that this matrix can be represented as a sum of the direct products of the density matrices for the individual spins. This means (now from the "first principles") that the state is separable, i.e., unentangled; $E(T_E) = 0$.

The entanglement of a system in a mixed triplet state is also zero (although one function among Eqs. (28)–(30), $|\psi_3\rangle$, is not factorizable). To verify this, let us pass to the limit $K \longrightarrow +\infty$ in Eq. (10). We obtain

$$\rho = \frac{1}{6}\begin{pmatrix} 2 & & & \\ & 1 & 1 & \\ & 1 & 1 & \\ & & & 2 \end{pmatrix} = \frac{1}{3}\sum_{i=1}^{3}|\psi_i\rangle\langle\psi_i|. \quad (34)$$

For this matrix, we again found a decomposition that explicitly demonstrates the separability of the triplet state (see Appendix).

As the result, the picture looks as follows. The entanglement of an antiferromagnetic dimer at $T = 0$ is unity (the maximally entangled state). As the temperature increases, the triplet is populated and the entanglement weakens. It disappears when the statistical weight of the singlet in the density matrix decreases to 1/2 (see (33)). Since there is the characteristic energy parameter $\Delta$ $(=2|J|)$ in the system, it is not surprising that the entanglement disappears at $T_E \sim \Delta/k_B$. Subsequently, at $T > T_E$, the system always remains in a separable state.

## 5. CONCLUSIONS

Based on a simple model of a spin dimer, we analyzed the experimental data on the magnetic susceptibility of paramagnetic nitrosyl iron complex $[Fe_2(SR)_2(NO)_4]$ with $R = C_3H_3N_2$ and $C_3H_5N_2$. Our analysis suggests that quantum-mechanical entanglement arises in both these compounds at nitrogen temperatures.

For the complex with $R = C_3H_5N_2$, we presented the temperature dependence of the entanglement. Experimental data show that the degree of entanglement in this NIC is close to 100% when the temperature decreases to $T = 25$ K.

A high temperature $T_E$ of the paramagnetic materials is their significant advantage over the materials with nuclear spins, where, as follows from estimates [31], entanglement can arise only at tenths of a microkelvin.

The so-called diamagnetic NICs [32] are of interest in that the temperature $T_E$ can be increased.

Their molecular structure is

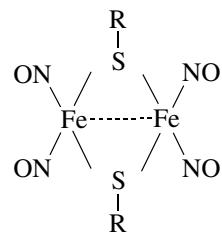

Here, R are now the six-membered rings $C_{6-n}H_{5-n}N_n$ ($n = 0, 1, 2$). In these NICs, the magnetoactive $Fe(NO)_2$ centers also form antiferromagnetic dimers. Since the ligands R in the molecule are directly connected only with sulfur, the separation between the iron atoms in such complexes is shortened to 0.27 nm. Obviously, a small separation causes a strong exchange interaction in the dimer. Therefore, we can assume that the state of the magnetic dimers in the material remains a singlet (and, hence, entangled) one even at room temperatures.

What are the interdimer interactions in NIC crystals is an important question. Their determination would open a possibility of investigating the entanglement on macroscopic scales (and not only inside molecules). More sophisticated experiments are required to study the weak interdimer couplings.

The method of investigating the entanglement developed here is general and applicable not only to NICs but also to other objects containing dimers.

## ACKNOWLEDGMENTS

We thank A.V. Kulikov, A.F. Shestakov, and S.I. Doronin for helpful discussions. We are also grateful to E.I. Kuznetsova who helped in preparing the manuscript. This work was supported by the Russian Foundation for Basic Research (project. no. 07-07-00048).

## APPENDIX

*Definition* (see [33, 34]). The state of a system that consists of two subsystems, 1 and 2, is called separable if there exists at least one decomposition of the system's density matrix $\rho$ in the form

$$\rho = \sum_i w_i \rho_i^{(1)} \otimes \rho_i^{(2)}, \quad (A.1)$$

where the weights $w_i \geq 0$ and $\sum_i w_i = 1$, while $\rho_i^{(1,2)}$ are the density matrices of subsystems 1 and 2.

If the system is separable, then there is no entanglement: $E = 0$. Conversely, inseparability means that there is entanglement in the system: $E \neq 0$.

We found through direct calculations that the density matrix (33) has the decomposition

$$\frac{1}{6}\begin{pmatrix} 1 & & & \\ & 2 & -1 & \\ & -1 & 2 & \\ & & & 1 \end{pmatrix} = \frac{1}{6}\left[\begin{pmatrix} 1 & 0 \\ 0 & 0 \end{pmatrix}\right.$$

$$\otimes \begin{pmatrix} 0 & 0 \\ 0 & 1 \end{pmatrix} + \begin{pmatrix} 0 & 0 \\ 0 & 1 \end{pmatrix} \otimes \begin{pmatrix} 1 & 0 \\ 0 & 0 \end{pmatrix}$$

$$+ \begin{pmatrix} \frac{1}{2} & \frac{1}{2} \\ \frac{1}{2} & \frac{1}{2} \end{pmatrix} \otimes \begin{pmatrix} \frac{1}{2} & -\frac{1}{2} \\ -\frac{1}{2} & \frac{1}{2} \end{pmatrix} + \begin{pmatrix} \frac{1}{2} & -\frac{1}{2} \\ -\frac{1}{2} & \frac{1}{2} \end{pmatrix}$$

$$\otimes \begin{pmatrix} \frac{1}{2} & \frac{1}{2} \\ \frac{1}{2} & \frac{1}{2} \end{pmatrix} + \begin{pmatrix} \frac{1}{2} & \frac{i}{2} \\ -\frac{i}{2} & \frac{1}{2} \end{pmatrix} \otimes \begin{pmatrix} \frac{1}{2} & -\frac{i}{2} \\ \frac{i}{2} & \frac{1}{2} \end{pmatrix}$$

$$+ \begin{pmatrix} \frac{1}{2} & -\frac{i}{2} \\ \frac{i}{2} & \frac{1}{2} \end{pmatrix} \otimes \begin{pmatrix} \frac{1}{2} & \frac{i}{2} \\ -\frac{i}{2} & \frac{1}{2} \end{pmatrix}\right].$$

Similarly, for the density matrix (34), we obtained

$$\frac{1}{6}\begin{pmatrix} 2 & & & \\ & 1 & 1 & \\ & 1 & 1 & \\ & & & 2 \end{pmatrix} = \frac{1}{6}\left[\begin{pmatrix} 1 & 0 \\ 0 & 0 \end{pmatrix}\right.$$

$$\otimes \begin{pmatrix} 1 & 0 \\ 0 & 0 \end{pmatrix} + \begin{pmatrix} 0 & 0 \\ 0 & 1 \end{pmatrix} \otimes \begin{pmatrix} 0 & 0 \\ 0 & 1 \end{pmatrix}$$

$$+ \begin{pmatrix} \frac{1}{2} & \frac{1}{2} \\ \frac{1}{2} & \frac{1}{2} \end{pmatrix} \otimes \begin{pmatrix} \frac{1}{2} & \frac{1}{2} \\ \frac{1}{2} & \frac{1}{2} \end{pmatrix} + \begin{pmatrix} \frac{1}{2} & -\frac{1}{2} \\ -\frac{1}{2} & \frac{1}{2} \end{pmatrix}$$

$$\otimes \begin{pmatrix} \frac{1}{2} & -\frac{1}{2} \\ -\frac{1}{2} & \frac{1}{2} \end{pmatrix} + \begin{pmatrix} \frac{1}{2} & \frac{i}{2} \\ -\frac{i}{2} & \frac{1}{2} \end{pmatrix} \otimes \begin{pmatrix} \frac{1}{2} & -\frac{i}{2} \\ \frac{i}{2} & \frac{1}{2} \end{pmatrix}$$

$$+ \begin{pmatrix} \frac{1}{2} & -\frac{i}{2} \\ \frac{i}{2} & \frac{1}{2} \end{pmatrix} \otimes \begin{pmatrix} \frac{1}{2} & -\frac{i}{2} \\ \frac{i}{2} & \frac{1}{2} \end{pmatrix}\right].$$

Both these decompositions satisfy Eq. (A.1). Therefore, the density matrices (33) and (34) correspond to separable states, i.e., $E = 0$ in these states.

## REFERENCES

1. V. V. Mityugov, Usp. Fiz. Nauk **163** (8), 103 (1993) [Phys.—Usp. **36** (8), 744 (1993)].
2. M. A. Nielsen and I. L. Chuang, *Quantum Computation and Quantum Information* (Cambridge University Press, Cambridge, 2000).
3. *The Physics of Quantum Information: Quantum Cryptography, Quantum Teleportation, and Quantum Computation*, Ed. by D. Bouwmeester, A. Ekert, and

A. Zeilinger (Springer, Berlin, 2000; Postmarket, Moscow, 2002).

4. A. A. Kokin, *Solid State Quantum Computers on Nuclear Spins* (Computer Science Institute, Russian Academy of Sciences, Moscow, 2004) [in Russian].

5. K. A. Valiev, Usp. Fiz. Nauk **175** (1), 3 (2005) [Phys.—Usp. **48** (1), 1 (2005)].

6. X. Wang and P. Zanardi, Phys. Lett. A **301**, 1 (2002).

7. R. A. Cowley, J. Phys.: Condens. Matter **15**, 4143 (2003).

8. Č. Brukner and V. Vedral, arXiv:quant-ph/0406040.

9. M. Wieśniak, V. Vedral, and Č. Brukner, New J. Phys. **7**, 258 (2005).

10. L. Amico, R. Fazio, A. Osterloh, and V. Vedral, Rev. Mod. Phys. **80**, 517 (2008); arXiv:quant-ph/0703044.

11. M. Wieśniak, *Quantum Entanglement in Some Physical Systems*, Dissertation (Gdańsk, 2007); arXiv:quant-ph/0710.1775.

12. Č. Brukner, V. Vedral, and A. Zeilinger, Phys. Rev. A: At., Mol., Opt. Phys. **73**, 012 110 (2006).

13. A. M. Souza, M. S. Reis, D. O. Soares-Pinto, et al., Phys. Rev. B: Condens. Matter **77**, 104 402 (2008).

14. N. A. Sanina, S. M. Aldoshin, T. N. Rudneva, et al., J. Mol. Struct. **752**, 110 (2005).

15. N. A. Sanina, T. N. Rudneva, S. M. Aldoshin, et al., Izv. Akad. Nauk, Ser. Khim., No. 1, 28 (2007).

16. A. F. Vanin, Usp. Fiz. Nauk **170** (4), 455 (2000) [Phys.—Usp. **43** (4), 415 (2000)].

17. N. A. Sanina and S. M. Aldoshin, Izv. Akad. Nauk, Ser. Khim., No. 11, 2326 (2004).

18. T. N. Rudneva, "*Synthesis and Investigation of the Structure and NO-Donor Activity of Nitrosyl Iron Complexes with 2-Mercaptoimidazoles*," Candidate's Dissertation in Chemistry (Institute of Problems of Chemical Physics, Russian Academy of Sciences, Chernogolovka, Moscow region, 2007); http://www.icp.ac.ru/news/avtoref/070628_Rudneva.doc.

19. A. F. Shestakov, Yu. M. Shul'ga, N. S. Emel'yanova, et al., Izv. Akad. Nauk, Ser. Khim., No. 7, 1244 (2007).

20. A. N. Vasil'ev, M. M. Markina, and E. A. Popova, Fiz. Nizk. Temp. (Kharkov) **31** (3), 272 (2005) [Low Temp. Phys. **31** (3), 203 (2005)].

21. A. I. Smirnov and V. N. Glazkov, Zh. Éksp. Teor. Fiz. **132** (4), 984 (2007) [JETP **105** (4), 861 (2007)].

22. B. Bleaney and K. D. Bowers, Proc. R. Soc. London, Ser. A **214**, 451 (1952).

23. R. L. Carlin, *Magnetochemistry* (Springer, New York, 1986; Mir, Moscow, 1989).

24. C. H. Bennett, D. P. DiVincenzo, J. A. Smolin, and W. K. Wootters, Phys. Rev. A: At., Mol., Opt. Phys. **54**, 3824 (1996).

25. S. Hill and W. K. Wootters, Phys. Rev. Lett. **78**, 5022 (1997).

26. W. K. Wootters, Phys. Rev. Lett. **80**, 2245 (1998).

27. K. M. O'Connor and W. K. Wootters, Phys. Rev. A: At., Mol., Opt. Phys. **63**, 052 302 (2002).

28. M. A. Nielsen, "Quantum Information Theory," Dissertation (New Mexico, 1998); arXiv:quant-ph/0011036.

29. M. C. Arnesen, S. Bose, and V. Vedral, Phys. Rev. Lett. **87**, 017 901 (2001).

30. A. E. Dubinov, I. D. Dubinova, and S. K. Saĭkov, *Lambert W-Function: Table of Integrals and Other Mathematical Properties* (Sarov Physicotechnical Institute, Sarov, 2004).

31. S. I. Doronin, A. N. Pyrkov, and É. B. Fel'dman, Pis'ma Zh. Éksp. Teor. Fiz. **85** (10), 627 (2007) [JETP Lett. **85** (10), 519 (2007)]; Zh. Éksp. Teor. Fiz. **132** (5), 1091 (2007) [JETP **105** (5), 953 (2007)].

32. A. F. Shestakov, Yu. M. Shul'ga, N. S. Emel'yanova, et al., Izv. Akad. Nauk, Ser. Khim., No. 12, 2053 (2006).

33. R. F. Werner, Phys. Rev. A: At., Mol., Opt. Phys. **40**, 4277 (1989).

34. A. Peres, Phys. Rev. Lett. **77**, 1413 (1996).